# Accurate prediction of nanovoid structures and energetics in bcc metals

Jie Hou[1,2], Yu-Wei You[1], Xiang-Shan Kong[1,3*], Jun Song[2*], C. S. Liu[1]

[1]Key Laboratory of Materials Physics, Institute of Solid State Physics, Chinese Academy of Sciences, P. O. Box 1129, Hefei 230031, P. R. China

[2]Department of Mining and Materials Engineering, McGill University, Montreal, Quebec H3A 0C5, Canada.

[3]Key Laboratory for Liquid-Solid Structural Evolution and Processing of Materials, Ministry of Education, Shandong University, Jinan, Shandong 250061, PR China

**Abstract**

Knowledge on structures and energetics of nanovoids is fundamental to understand defect evolution in metals. Yet there remain no reliable methods able to determine essential structural details or to provide accurate assessment of energetics for general nanovoids. Here, we performed systematic first-principles investigations to examine stable structures and energetics of nanovoids in bcc metals, explicitly demonstrated the stable structures can be precisely determined by minimizing their Wigner-Seitz area, and revealed a linear relationship between formation energy and Wigner-Seitz area of nanovoids. We further developed a new physics-based model to accurately predict stable structures and energetics for arbitrary-sized nanovoids. This model was well validated by first-principles calculations and recent nanovoid annealing experiments, and showed distinct advantages over the widely used spherical approximation. The present work offers mechanistic insights that crucial for understanding nanovoid formation and evolution, being a critical step towards predictive control and prevention of nanovoid related damage processes in structural metals.

**Introduction**

Nanometer-sized voids, or namely nanovoids, one of most common defect groups present in metals under deformation [1,2], irradiation [3-6], or after quenching [7,8], play a crucial role in affecting or determining many material properties and performance in various applications. Abundant nanovoids may develop into high density of cavities to subsequently give rise to volumetric swelling in nuclear materials [4,5,9-11], cause fracture in load-bearing components [2,12], and promote gaseous bubble formation in plasma-facing or cathodic protected metals [13-16]. These nanovoid-related features greatly alter the properties and performance of materials and thus necessitate thorough studies on the formation and evolution of nanovoids.

Despite decades of extensive research efforts, precise determination of stable structures and energetics of nanovoids remain a formidable challenge. The small-scale details of nanovoids are generally beyond the reach of experimentation and necessitate atomistic simulation techniques. First-principles density functional theory (DFT) calculations [17, 18] provide an accurate method to examine atomic details, and thus have been employed in numerous studies on nanovoids. For instance, as a key part of the tool chain for obtaining multiscale insights, DFT calculations were widely applied to study nanovoid formation and growth in nuclear materials such as W and α-Fe [13, 19-25]. Nonetheless, lacking knowledge of energy-structure relationship of nanovoids, those brute-force DFT explorations do not provide a systematic method in search of stable nanovoid structures, and are limited to small nanovoids containing less than 10 vacancies [13, 19-23] or very large voids which can be viewed as a set of flat surfaces [24, 25]. In light of such limitation, large-scale simulations based on empirical interatomic potentials were also performed to assist the structure search of nanovoids [26-28]. However, these simulations are also *ad hoc* in nature, therefore cannot overrun the exponential growth in number of possible structures as nanovoid size increases [29]. Precise determination of stable structures and energies is still limited to sizes around tens of vacancies even assisted by metaheuristic algorithms [28]. More importantly, most empirical potentials suffer from inadequate accuracy in their description of structural characteristics and energetics of nanovoid surfaces [26, 30]. For instance, even for the simplest case of divacancies in W, empirical potentials generally predict a strong binding between vacancies, yet DFT studies indicate a weak repulsion [21, 22, 31, 32]. Such inaccuracy will likely extend to general sized nanovoids, rendering the reliability of structures and energetics obtained on base of empirical potentials questionable. Therefore, there remains no computational pathway capable of accurate assessment of thermodynamics and structures of arbitrary-sized nanovoids, which in turn results in deficit in the information required for multiscale studies.

Besides the atomistic investigation, an alternative approach often used to characterize general nanovoids is to assume they have spherical geometry with formation energies in linear relation to their surface area (often referred to as capillary law) [22, 33, 34]. Being easy to implement, able to capture the average trend of binding energy and without the size constrain in nanovoids, the spherical approximation have seen wide use in dozens of kinetic Monte Carlo or mean-field studies to parameterize nanovoids [13, 19, 22, 35-37]. Nonetheless, the spherical approximation inherently smears out structural details that are usually vital for evaluating energies of nanosized clusters [38, 39], therefore fails to reproduce the large binding energy fluctuation revealed by atomistic calculations [27, 34]. Such intrinsic inaccuracy in energetic and structural evaluations of nanovoids will necessarily cast doubts in its predictions of nanovoid formation and growth, and subsequent understanding of nanovoid mediated phenomena such as solute precipitation, void swelling, and bubble formation.

To address the afore-mentioned challenges, the present study aims to develop a reliable theoretical approach capable to provide accurate assessment of energetics and structures of general nanovoids. The focus is placed on bcc metals given their wide applications in nuclear industry where abundant nanovoids will be induced by irradiation during service. Starting with comprehensive DFT calculations, the stable structures of and energetics of nanovoids were systematically explored computationally. Based on the results, we explicitly demonstrated that stable structures of nanovoids can be precisely determined by minimizing their Wigner-Seitz areas, and revealed a linear relationship between the formation energy and Wigner-Seitz area for the nanovoids. We then established a physical model based on the structure-energy relationships found. This new model, validated by DFT calculations and recent nanovoid annealing experiments, provides a full-scale, accurate formation and binding energetics prediction for arbitrary-sized nanovoids, offering critical new knowledge and tools in understanding and analyzing nanovoid formation and evolution in structural metals.

**Results and Discussion**

**Stable structures and formation energies of nanovoids.** We started by performing a comprehensive set of DFT calculations screening different structures of general nanovoids $V_n$ with $n$ spanning from 1 to 7, in the four representative metal systems W, Mo, Ta and $\alpha$-Fe. Here the symbol $V_n$ indicates that the nanovoid is a product of $n$ vacancies coalescence. Multiple nanovoid structures were generated with the structure fulfilling the linkage constrain, i.e., each constituting vacancy maintain contact (1$^{st}$ or 2$^{nd}$ nearest neighbor) with at least one other vacancy in the nanovoid. For each structure generated, DFT calculations were performed to relax the structure and examine its formation energy:

$$E_f^{V_n} = E_{tot}^{V_n} - \frac{N-n}{N} E_{tot}^{bulk},$$

where $E_{tot}^{V_n}$ and $E_{tot}^{bulk}$ respectively represents the total energy of a supercell with and without the $V_n$ nanovoid, and $N$ is the number of metal atoms in the corresponding perfect supercell.

It is generally believed that compact structures are more favorable than linear or planar ones for nanovoids containing up to thousands of vacancies [26, 40, 41]. Here, we found that the Wigner-Seitz area of nanovoids serves to quantify the compactness and stability of different structures. The Wigner-Seitz area refers to the external surface area of Wigner-Seitz polyhedrons of vacancies constituting the nanovoid (note internal Wigner-Seitz boundaries between vacancies do not contribute to this area). Fig. 1 presents formation energies of nanovoids in different metals as functions of their respective Wigner-Seitz areas. We can see that for a particular nanovoid size, the nanovoid of the minimum Wigner-Seitz area always exhibit the lowest formation energy. The only exception is $V_2$ in Ta and α-Fe, where the compact structures have formation energy higher than (but close to) the loose structures (similar results were also found and discussed in Ref. [23]). This suggests that the most stable structure of a $E_f^{V_n}$ nanovoid be the one with the smallest Wigner-Seitz area, and we may determine its configuration by geometrically minimizing the Wigner-Seitz area without the need of time-consuming DFT structure search. Meanwhile, more intriguingly from Fig. 1 is that the formation energies of the most stable structures show a nice linear scaling with the Wigner-Seitz area, highlighting the possibility of predicting energetics of stable nanovoids simply based on their Wigner-Seitz area.

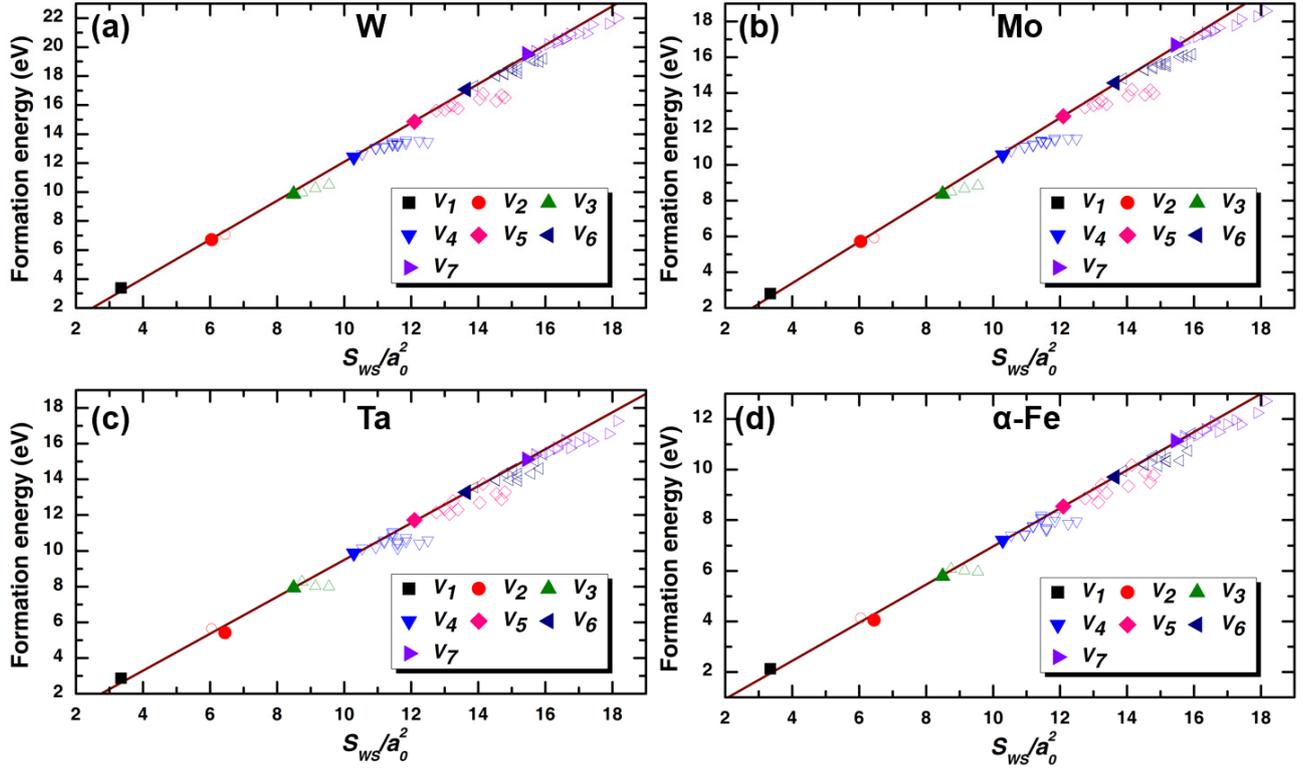

**Figure 1 | Formation energies of $V_1$-$V_7$ nanovoids in W (a), Mo (b), Ta (c) and α-Fe (d)** as functions of the Wigner-Seitz area of a nanovoid, $S_{WS}$ normalized by $a_0^2$, where $a_0$ denotes the lattice constant of the related metal. Solid symbols highlight the most stable structures with lines show corresponding linear fittings. Hollow symbols represent metastable structures.

To further verify the linear relationship observed in Fig. 1, we extend our investigation to examine larger nanovoids $V_n$ with *n* up to 20. As previously suggested by the results in Fig. 1, for each size *n*, the most stable nanovoid configuration, i.e., the one of the lowest formation energy, would correspond to the one with the lowest Wigner-Seitz area. Such configurations were consequently identified by minimizing the Wigner-Seitz area via simulated annealing algorithm (see Fig. 2a and Supplementary Materials for detailed structures), and additional benchmark calculations have been performed to verify that they are indeed the most stable nanovoid structures (see Figs. S1 and S4 in Supplementary Materials). Their formation energies versus their respective Wigner-Seitz area are presented in Fig. 2b. We see that all data consolidate perfectly into a single linear dependence. Based on the above, we can evaluate a stable nanovoid's formation energy simply based on its Wigner-Seitz area, i.e.:

$$E_f^{V_n} = A + C \cdot S_{WS}^{V_n}, \qquad (1)$$

where $S_{WS}^{V_n}$ is the Wigner-Seitz area of $V_n$ nanovoid, and $A$ and $C$ are two material constants. Benchmark simulations show that this linear relation prevails with further increase in the nanovoid size up to $V_{47}$, and the key parameter $C$ (i.e., line slopes in Fig. 1 and Fig. 2b.) is not sensitive to nanovoid size (see Fig. S2 and Table S1 in Supplementary Materials). This also suggest that Eq. (1) be valid for arbitrary sized nanovoids.

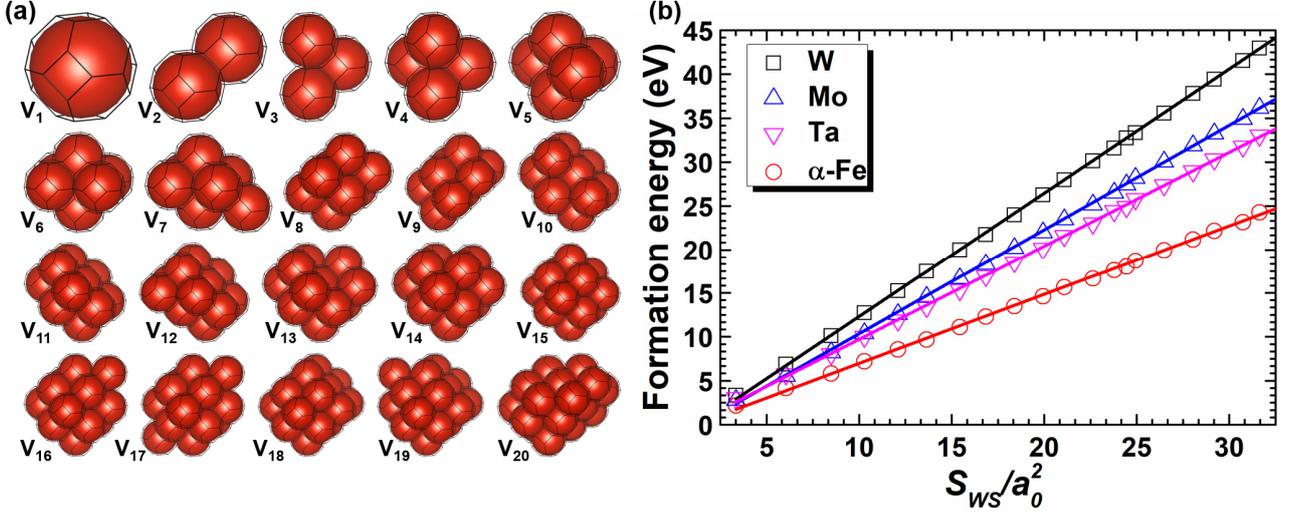

**Figure 2 | Structures and energetics of stable $V_1$-$V_{20}$ nanovoids in different metals.** (a) Atomic structures of these nanovoids determined by minimizing Wigner-Seitz area, where red spheres indicate constituting vacancies and black lines show edges of Wigner-Seitz cells on the nanovoid surface. (b) Formation energies as functions of the Wigner-Seitz area ($S_{WS}$) normalized by $a_0^2$ with $a_0$ being the lattice constant of the metal, where the solid lines indicate the corresponding linear fitting.

**Prediction of vacancy binding energy.** One metric of great importance in the evolution of nanovoid is the binding energy between a vacancy and nanovoid, which can be readily calculated from the nanovoid formation energy:

$$E_b^{V_n} = E_f^{V_1} + E_f^{V_{n-1}} - E_f^{V_n}. \tag{2}$$

In the above, $E_b^{V_n}$ denotes the binding energy of a vacancy in a nanovoid $V_n$, effectively representing the energy required to strip a vacancy away from a $V_n$ nanovoid. Based on the linear relationship conveyed by Eq. (1), the subtraction term in Eq. (2), i.e., $E_f^{V_{n-1}} - E_f^{V_n}$, should be proportional to the difference in Wigner-Seitz area between $V_n$ and $V_{n-1}$, i.e., $\Delta S_{WS}^{V_n} = S_{WS}^{V_n} - S_{WS}^{V_{n-1}}$. In this way, we can rewrite Eq. (2) into:

$$E_b^{V_n} = E_f^{V_1} - C\Delta S_{WS}^{V_n}.$$

Since $C$ is not sensitive to nanovoid size, it can be well estimated based on formation energies of two small nanovoids such as $V_1$ and $V_2$, i.e., $C = (E_f^{V_2} - E_f^{V_1})/\Delta S_{WS}^{V_2}$. Therefore, we have:

$$E_b^{V_n} = E_f^{V_1} - \left(E_f^{V_2} - E_f^{V_1}\right)\frac{\Delta S_{WS}^{V_n}}{\Delta S_{WS}^{V_2}}. \tag{3}$$

By employing Eq. (3), we are able to predict the binding energy of an arbitrary-sized nanovoid knowing just the formation energies of single vacancy and divacancy (similar results were found using best fitted $C$, further confirming that $C$ is not sensitive to nanovoid size, see Fig. S3 in Supplementary Materials). Fig. 3 presents both the predicted (from Eq. (3)) and DFT-calculated binding energy values, plotted as functions of the Wigner-Seitz area difference $\Delta S_{WS}^{V_n}$ (normalized by $a_0^2$). We see that the binding energy exhibits a linear decrease as $\Delta S_{WS}^{V_n}$ increases, as to be expected from Eq. (3). Meanwhile, an excellent agreement between the model predictions and DFT calculations is observed, evidencing the validity of our model.

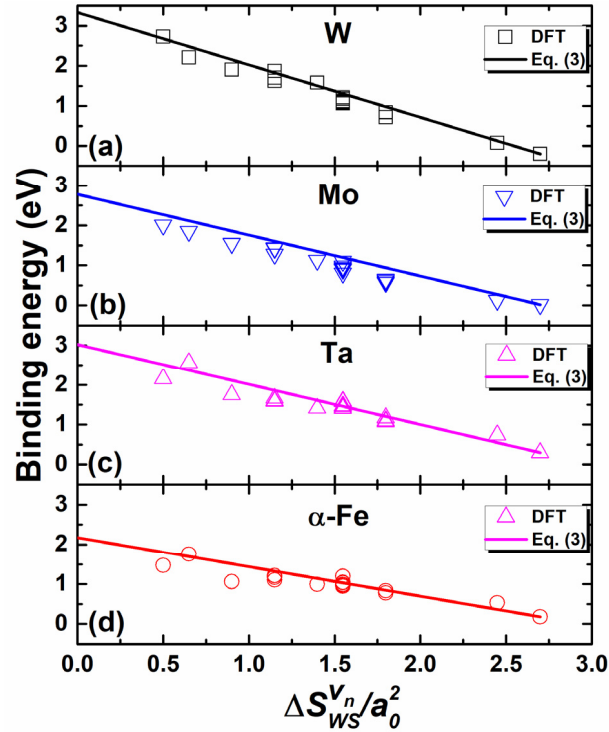

**Figure 3 | Vacancy binding energies for $V_2$-$V_{20}$ nanovoids in W (a), Mo (b), Ta (c) and α-Fe (d),** as functions of the Wigner-Seitz area change ($\Delta S_{WS}^{V_n}$) between $V_n$ and $V_{n-1}$ nanovoids normalized by $a_0^2$ with $a_0$ being the lattice constant of the metal. Symbols are DFT results. Lines are predictions provided by Eq. (3).

In Fig. 4, we further plotted the binding energy $E_b^{V_n}$ and Wigner-Seitz area difference $\Delta S_{WS}^{V_n}$ as functions of the nanovoid size $n$. Here, DFT calculations were extended to larger nanovoids (up to $V_{47}$) using W as a representative. Overall, as the size increases, the DFT-calculated binding energy increases, which can be attributed to decreasing in its area-to-size ratio that results in lower $\Delta S_{WS}^{V_n}$ as a nanovoid grows. Besides this general tendency, the evolution of both $E_b^{V_n}$ and $\Delta S_{WS}^{V_n}$ is not monotonic, but exhibits an irregular sawtooth-like pattern, with abrupt fluctuation in their values over the course of the evolution. Along with the DFT data, the binding energies predicted by Eq. (3) are also presented in Fig. 4a, showing excellent agreement with both the overall increasing trend and abrupt local variation in binding energy reproduced.

The fluctuations in binding energy come from vacancy binding with different sites on nanovoid surface. In analogy to the classical terrace-ledge-kink model of crystal growth [42], large nanovoid will provide more kink-like sites on its surface. Vacancy binding on these kink-like sites will induce little or no change in Wigner-Seitz area, i.e., a small $\Delta S_{WS}^{V_n}$, thus yield a high binding energy close or equal to $E_f^{V_1}$ according to Eq. (3). However, when a nanovoid reaches a closure of its outer shell, further binding of vacancy will occur at terrace-like or ledge-like sites, creating substantial new Wigner-Seitz area to the surface (large $\Delta S_{WS}^{V_n}$) and lower the binding energy accordingly. An example is given by $V_{15}$ and $V_{16}$ nanovoids. As demonstrated in Fig 4b, a $V_{15}$ nanovoid show close-shelled rhombic dodecahedron shape which is highly compact in Wigner-Seitz area, therefore has a relatively low $\Delta S_{WS}^{V_{15}}$ (0.5 $a_0^2$) and high $E_b^{V_{15}}$ (2.47 eV) in reference to $V_{14}$. While further vacancy binding can only occur at terrace-like sites, result in a higher $\Delta S_{WS}^{V_{16}}$ (1.55 $a_0^2$) and lower $E_b^{V_{16}}$ (0.88 eV).

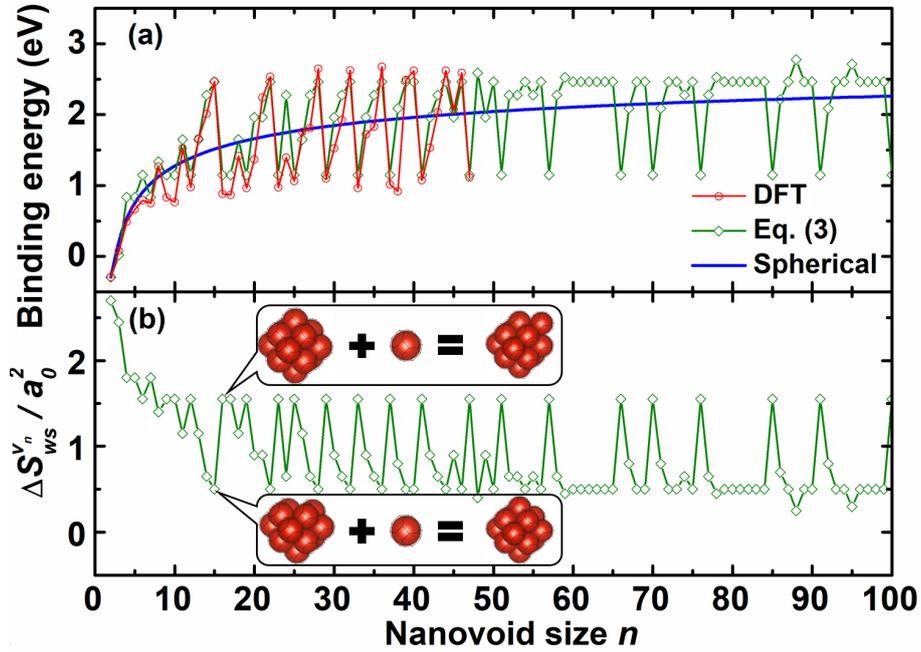

**Figure 4 | Energy and area changes during vacancy binding.** (a) Binding energies by DFT calculation and by different model predictions for nanovoids in W. (b) Wigner-Seitz area change ($\Delta S_{WS}^{V_n}$) for the most compact nanovoids in bcc metals normalized by $a_0^2$ with $a_0$ being the lattice constant of the metal.

**Comparison with spherical approximation and experiments.** As previously mentioned, the spherical approximation, which assumes nanovoids to be spherical in shape, has been a commonly used means to predict the energetics of nanovoids [13, 19, 22, 35, 36]. Binding energies predicted by the spherical approximation method are shown in Fig. 4a (cf. solid blue curve). Apparently, the spherical approximation is not able to capture the large fluctuation in the binding energy evolution revealed by our model (Eq. (3)) and DFT calculations. Despite its ability to recognize the general, average trend, the spherical approximation fails to provide accurate prediction of the binding energies, which consequently would lead to erroneous evaluation of thermal stabilities of nanovoids.

Such deficit in the spherical approximation is clearly evident when examining the average lifetime of nanovoids, i.e., the time required for a complete dissociation of nanovoids via vacancy thermal emission. The emission rate is sensitive to the emission barrier of vacancy, which approximately equals to the sum of nanovoid's vacancy binding energy and vacancy migration barrier in bulk (1.66 eV in W by Ref. [22]). Fig. 5a shows the average lifetime of nanovoids predicted by object kinetic Monte Carlo (OKMC) simulations, with nanovoid energetics inputs separately from DFT calculations, our model and the spherical approximation. As seen from Fig. 5a, despite similar lifetime prediction for small nanovoids (i.e., $V_n$ with $n<12$, the size range most previous DFT studies [13, 21, 22, 26, 40] focused on), the spherical approximation significantly underestimates the lifetime by orders of magnitude for larger nanovoids. In contrast, the lifetime predicted based on our model is in close agreement with the one based on DFT calculations throughout the nanovoid size range explored.

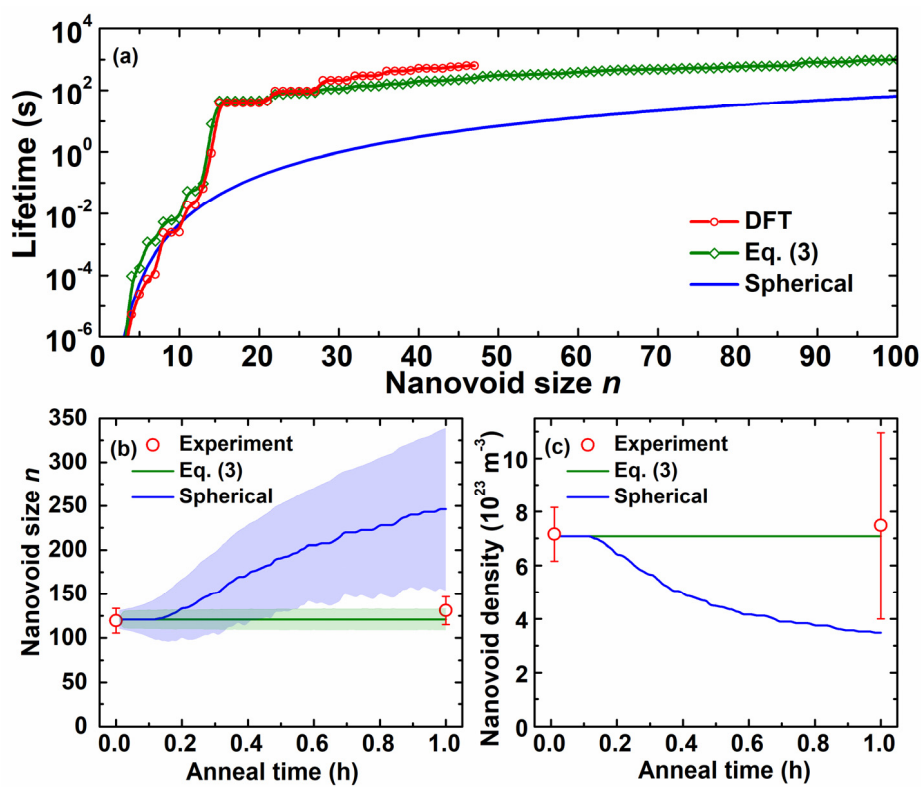

**Figure 5 | OKMC simulation results with different binding energy inputs from Fig. 4a.** (a) Average lifetime

of nanovoids against thermal dissociation at 1373 K. (b) Average nanovoid size, and (c) nanovoid density as functions of anneal time at 1373 K. Shaded area and bars are corresponding standard derivations. Experimental results were extracted from Ref. [5].

Furthermore, our model is also supported by recent post-irradiation annealing experiment results by Ferroni et al.[5], as elaborated in the follows. In their experimental study, 2 MeV self-ion irradiated tungsten sample was first annealed at low temperatures to create TEM-visible nanovoids with diameter around 1.54 nm (correspond to the size of $V_{120}$). Subsequently the sample was further annealed at 1373 K for an hour with neither size nor density of nanovoids found changed, suggesting that these nanovoids remain stable at 1373 K. In accordance to the experiment, corresponding OKMC simulations were performed, with results demonstrated in Fig. 5b and 5c. From the figures we note the conventional spherical approximation predicts an evident growth in nanovoid size along with a decrease in nanovoid density, showing clear deviations from the experimental results. These deviations directly originate from it underestimating thermal stability of nanovoids (cf. Fig. 5a). On the other hand, OKMC simulations based on our model well conform with the experimental observations.

The above good agreements between our model and DFT/experimental results clearly evidence that this newly proposed model captures the underlying physics of nanovoid formation and vacancy binding. This new model is physics-based with no empirical parameters, capable of quick determination of stable structures and accurate prediction of binding energies for arbitrary-sized nanovoids in bcc metals. It provides accurate atomistic data that are beyond the reach of direct DFT calculations, enabling multiscale modeling with coarse-grain methods such as OKMC or cluster dynamics simulations. Furthermore, the model may serve as critical benchmarks for evaluating and/or developing new empirical interatomic potentials. Here, representative Molecular Statics calculations [43] were performed using large super-cells with different potentials [26, 30, 31, 44] for W (see Figs. S4-S6 in Supplementary Materials for details), showing that empirical potentials reproduce the linear relationship (between formation energy and Wigner-Seitz area) found above, but yield large deviations in the values of formation energy and generally overestimate the binding energy. These efforts are expected to facilitate more reliable molecular dynamics simulations of nanovoid evolution in the future.

To sum up, the stable structures and energetics of nanovoids in bcc metals have been investigated by comprehensive DFT calculations, and a new physics-based predictive model has been developed for arbitrary-sized nanovoids. We demonstrated that stable structures of nanovoid can be precisely determined by minimizing their Wigner-Seitz surface area, and revealed a linear relationship between the formation energy and Wigner-Seitz area for these nanovoids. The new physical model developed captures the structure-energy relationships found, capable to provide a full-scale formation and binding energetics prediction for nanovoids. The new model, generally applicable for arbitrary-sized nanovoids in bcc metals, exhibits superb advantages over the widely used spherical approximation of nanovoids, not only offering significantly more accurate predictions but also recognizing atomic-scale structural variation in nanovoid evolution. The accuracy and reliability of our model is well validated by excellent consistency with DFT results and agreement with recent nanovoid annealing experiments. The present study provides critical new knowledge in understanding nanovoid formation and evolution, and sets a key step towards predictive control and prevention of nanovoid related damage processes in structural metals.

**Methods**

**First-principles density functional theory (DFT) calculations.** First-principles DFT calculations were performed employing the Vienna *ab initio* simulation package (VASP) [45, 46] with Blochl's projector augmented wave (PAW) potential method [47]. All the outer *d*-shell and *s*-shell electrons of metal were treated as valence electrons (3$d$ and 4$s$ for Fe, 4$d$ and 5$s$ for Mo, 5$d$ and 6$s$ for Ta and W). The exchange-correlation energy functional was described with the generalized gradient approximation (GGA) as parameterized by Perdew-Burke-Ernzerhof (PBE)[48]. For computational efficiency reasons, three sets of super-cells containing 128, 250, and 686 lattice points ($4 \times 4 \times 4$, $5 \times 5 \times 5$, and $7 \times 7 \times 7$ duplicates of a conventional bcc unit cell) were respectively used for calculations of $V_1$-$V_7$, $V_1$-$V_{20}$ and $V_1$-$V_{47}$ (i.e., data presented in Fig. 1, Figs. 2-3, and Fig. 4, respectively). A 500 eV plane wave cutoff was adopted for calculations using the $4 \times 4 \times 4$ and $5 \times 5 \times 5$ super-cells, while a 350 eV cutoff was used for the $7 \times 7 \times 7$ super-cell. A $3 \times 3 \times 3$ *k*-point grid obtained using the Monkhorst-Pack method [45] were used for all calculations. Partial occupancies were determined using first order Methfessel-Paxton method with a smearing width of 0.2 eV. Relaxations of atomic configuration and optimizations of the shape and size of the super-cell were performed for all calculations, with convergence criteria for energy and atomic force were set as $10^{-6}$ eV and 0.01 eV/Å respectively. Benchmark tests with above different settings show negligible influences on the results (see Fig. S2 in Supplementary Materials).

**Object kinetic Monte Carlo (OKMC) simulations.** OKMC simulations (see Ref. [22] for general algorithms of the method) were conducted to model thermal dissociation and annealing of nanovoids. In the thermal dissociation simulations, a nanovoid is placed in an infinite-large simulation box and can emit monovacancies until its complete dissociation. The monovacancy is removed from the system right after the emission with no chance of recombination. 100,000 simulations were conducted for each size of nanovoid to obtain a statistically converged average lifetime. In the annealing simulations, a $60 \times 60 \times 60$ nm$^3$ simulation box with periodic boundary condition applied was used to model bulk environments. A initial nanovoid size of $V_{120}$ and density of $7.16 \times 10^{23}$ $m^{-3}$ was adopted according to the related experimental observations [5]. Binding energies of nanovoids were parameterized according to Eq. (3) or spherical approximation as discussed in the main text. Other parameters, including migration barriers and attempt frequencies, were parameterized according to Ref. [22].

**Supplementary Materials**

Figure S1. Formation energies of $V_{15}$ nanovoids with different structures in different metals.

Figure S2. Formation energies as functions of the Wigner-Seitz area ($S_{WS}^{V_n}$) in different metals.

Table S1. Values of constant $C$ (eV/$a_0^2$) fitted using formation energies of different sizes of nanovoids with different super-cells.

Figure S3. Binding energies for $V_2$-$V_{20}$ nanovoids in different metals.

Figure S4. Formation energies of $V_{15}$ nanovoids with different structures in W based on different empirical potentials.

Figure S5. Formation energies of nanovoids in W by DFT and Molecular Statics calculation.

Figure S6. Binding energies of nanovoids by DFT calculation, model prediction, and Molecular statics calculation.

Table S2. Minimum Wigner-Seitz area of different $V_n$ nanovoids.

References ([26, 30, 31, 43, 44]).

Data files containing predicted nanovoid structures.

**Acknowledgement**

The authors are grateful for valuable comments and discussion from G.H. Lu at Beihang University and F. Gao at University of Michigan, Y.G. Li and Z.M. Xie at Institute of Solid State Physics, Chinese Academy of Sciences, J.T. Zhao and X. Meng at Lanzhou University. **Funding:** This work was financially supported by the National Key R&D Program of China (grant no. 2018YFE0308102), the National Natural Science Foundation of China (Nos.:11735015, 51771185, 11575229), Natural Sciences and Engineering Research Council of Canada grants (# RGPIN-2017-05187 and # RGPAS 507979-17). J. Hou and J. Song acknowledge Compute Canada and the Supercomputer Consortium Laval UQAM McGill and Eastern Quebec for providing computing resources. **Author contributions:** X.S.K. and J.S. designed the project. J.H. developed the OKMC model, J.H., X.S.K, and Y.W.Y conducted the simulations with help from C.S.L and J.S. All authors wrote the paper and discussed the results. C.S.L. and J.S. supervised the work. **Competing interests:** The authors declare no competing interests. **Data and code availability:** The data generated and/or analysed within the current study, and the code for the object kinetic Monte Carlo simulations, will be made available upon reasonable request to the authors.

# Supplementary materials for accurate prediction of nanovoid structures and energetics in bcc metals

Jie Hou[1,2], Yu-Wei You[1], Xiang-Shan Kong[1,3*], Jun Song[2*], C. S. Liu[1]

[1]Key Laboratory of Materials Physics, Institute of Solid State Physics, Chinese Academy of Sciences, P. O. Box 1129, Hefei 230031, P. R. China

[2]Department of Mining and Materials Engineering, McGill University, Montreal, Quebec H3A 0C5, Canada.

[3]Key Laboratory for Liquid-Solid Structural Evolution and Processing of Materials, Ministry of Education, Shandong University, Jinan, Shandong 250061, PR China

**This PDF file includes:**

Figure S1. Formation energies of $V_{15}$ nanovoids with different structures in different metals.

Figure S2. Formation energies as functions of the Wigner-Seitz area ($S_{WS}^{V_n}$) in different metals.

Table S1. Values of constant $C$ (eV/$a_0^2$) fitted using formation energies of different sizes of nanovoids with different super-cells.

Figure S3. Binding energies for $V_2$-$V_{20}$ nanovoids in different metals.

Figure S4. Formation energies of $V_{15}$ nanovoids with different structures in W based on different empirical potentials.

Figure S5. Formation energies of nanovoids in W by DFT and Molecular Statics calculation.

Figure S6. Binding energies of nanovoids by DFT calculation, model prediction, and Molecular statics calculation.

Table S2. Minimum Wigner-Seitz area of different $V_n$ nanovoids.

References (26, 30, 31, 43, 44).

---

* Correspondence should be addressed to X.S.K. (email: xskong@sdu.edu.cn) or to J.S. (email: jun.song2@mcgill.ca)

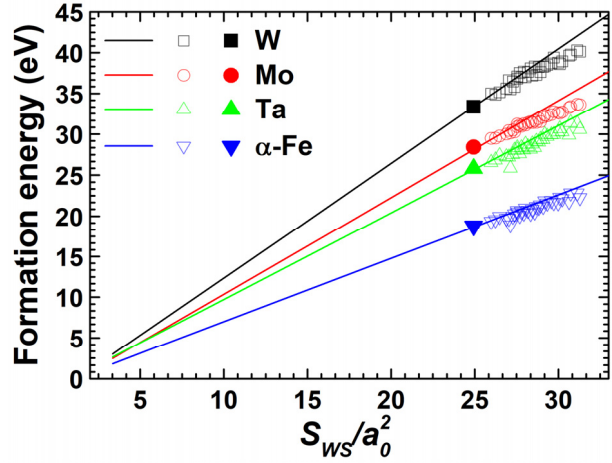

**Figure S1. Formation energies of $V_{15}$ nanovoids with different structures in different metals**, as functions of the Wigner-Seitz area of nanovoids ($S_{WS}$) normalized by $a_0^2$ with $a_0$ being the lattice constant. Solid symbols highlight the most stable structures with lines show corresponding linear fittings from cf. Fig. 2b. Hollow symbols represent metastable structures. Results were calculated using 5×5×5 super-cells.

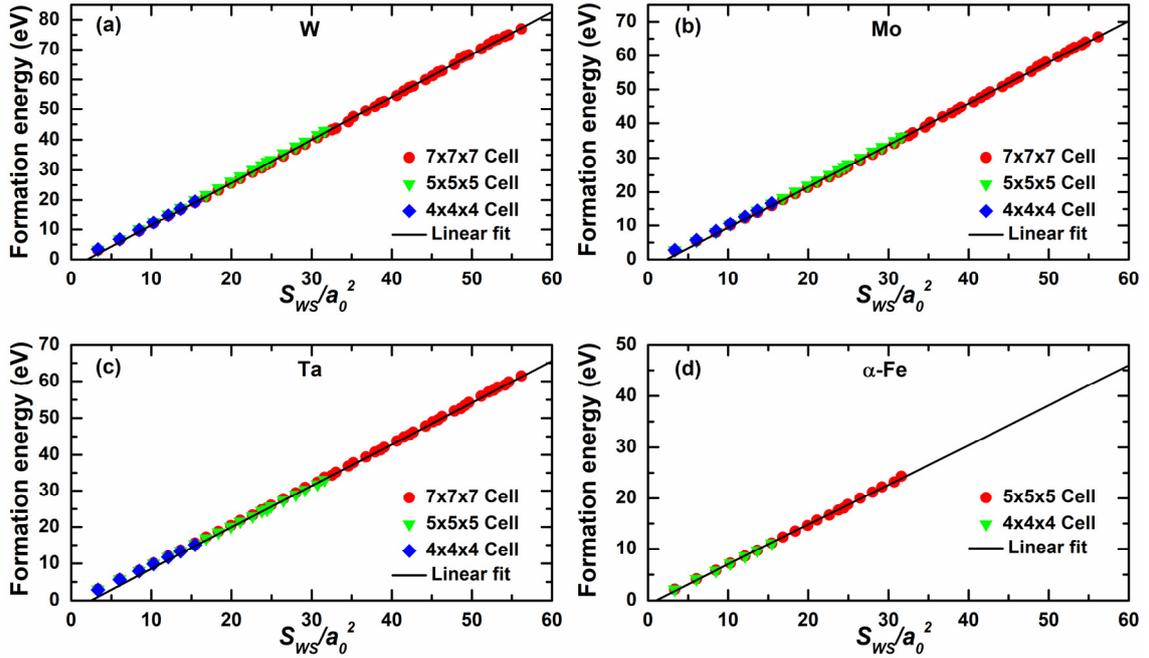

**Figure S2**. Formation energies as functions of the Wigner-Seitz area ($S_{WS}^{V_n}$) in different metals, normalized by $a_0^2$ with $a_0$ being the lattice constant of the metal. Dots show DFT calculated data for $V_1 - V_{47}$ nanovoids using different super-cells, solid lines indicate the corresponding linear fittings. Results for Fe are partially included here due to computational difficulties.

**Table S1**. **Values of constant $C$ (eV/$a_0^2$) fitted using formation energies of different sizes of nanovoids with different super-cells.** Results for α-Fe are partially included here due to computational difficulties.

| | $V_1 - V_7$ | $V_1 - V_{20}$ | $V_1 - V_{47}$ |
|---|---|---|---|
| | | | |

|  | 4×4×4 super-cell | 5×5×5 super-cell | 7×7×7 super-cell |
|---|---|---|---|
| **W** | 1.34 | 1.41 | 1.42 |
| **Mo** | 1.15 | 1.19 | 1.22 |
| **Ta** | 1.01 | 1.07 | 1.12 |
| **α-Fe** | 0.74 | 0.78 | |

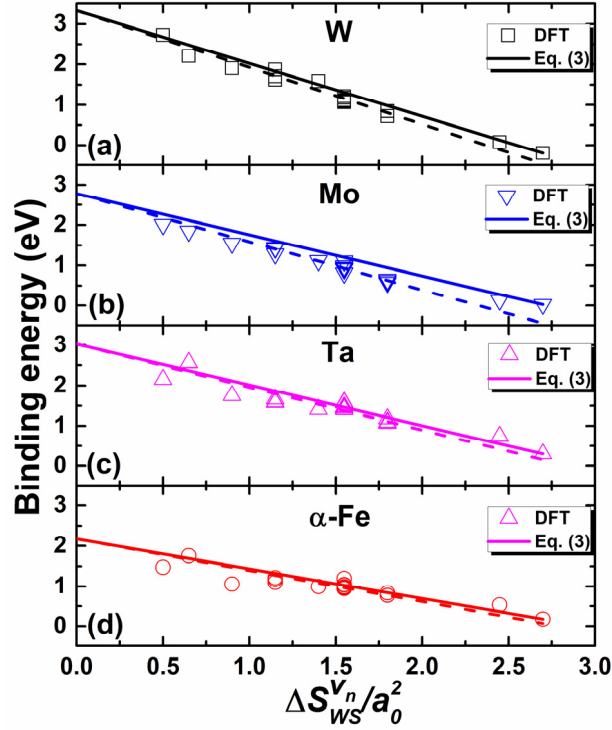

**Figure S3**. **Binding energies for $V_2$-$V_{20}$ nanovoids in different metals**, as functions of the Wigner-Seitz area change ($\Delta S_{WS}^{V_n}$ normalized by $a_0^2$ with $a_0$ being the lattice constant) between $V_n$ and $V_{n-1}$ nanovoids. Symbols are DFT results. Solid lines are our model predictions with $C = (E_f^{V_2} - E_f^{V_1})/\Delta S_{WS}^{V_2}$, (i.e., cf. Eq. (3)). Dashed lines are that with constant C fitted using formations energies of $V_1$-$V_{20}$.

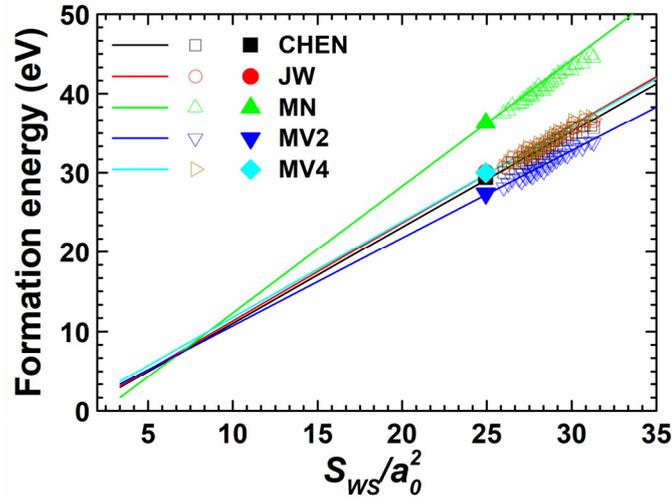

**Figure S4. Formation energies of $V_{15}$ nanovoids with different structures in W based on different empirical potentials**, as functions of the Wigner-Seitz area of nanovoids ($S_{WS}$) normalized by $a_0^2$ with $a_0$ being the lattice constant. Solid symbols highlight the most stable structures with lines show corresponding linear fittings from Figure S5. Hollow symbols represent metastable structures. Results are calculated using Molecular Statics based on different EAM empirical potentials that parameterized in ref.[30] (CHEN), ref.[44] (JW), ref.[26] (MN), ref.[31] (MV2 and MV4). All Molecular Statics calculations in this work were performed using the LAMMPS package[44] with a 50×50×50 super-cell. Relaxations of atomic configuration and optimizations of the shape and size of the super-cell were performed for all calculations.

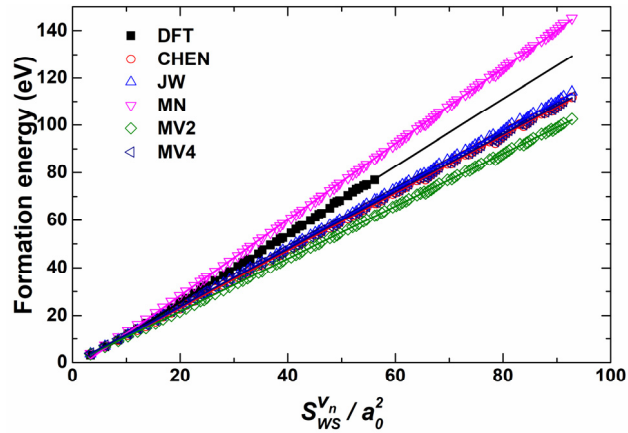

**Figure S5. Formation energies of nanovoids in W by DFT and Molecular Statics calculation**, as functions of the Wigner-Seitz area ($S_{WS}^{V_n}$) normalized by $a_0^2$ with $a_0$ being the lattice constant. Solid symbols are results from DFT calculations, hollow symbols are Molecular Statics results with same notation as in Figure S4. Note all EAM potentials reproduce the linear relationship, but show large variations in line slope.

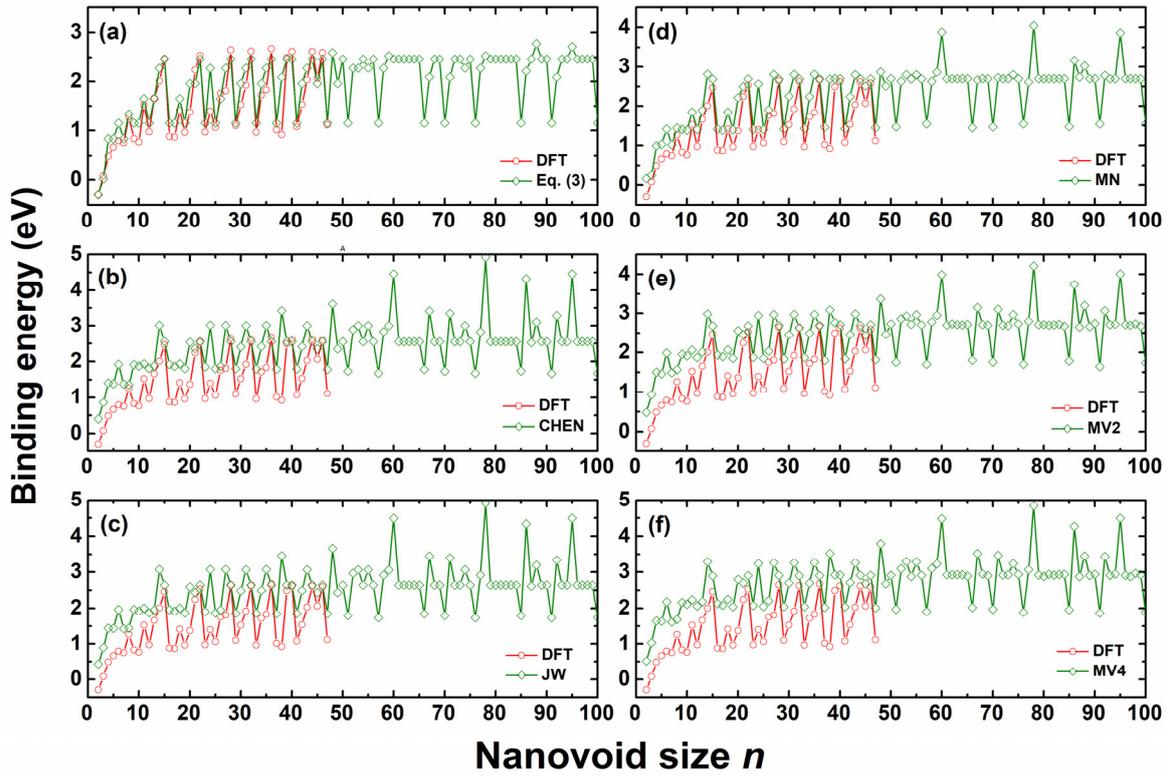

**Figure S6. Binding energies of nanovoids by DFT calculation, model prediction, and Molecular statics calculation.** (a) DFT results vs. cf. Eq. (3) prediction, (b-e) DFT results vs. Molecular Statics results based on different EAM potentials, same notation as in Figure S4 is used here. We can see that cf. Eq. (3) provides the most accurate predictions to DFT results. While among the EAM potentials tested here, the recently developed MN potential show better agreement with DFT results, but still overestimate the binding energy in general.

**Table S2. Minimum Wigner-Seitz area of different V$_n$ nanovoids,** i.e., $S_{WS}^{V_n}$ (normalized by $a_0^2$ with $a_0$ being the lattice constant), in bcc metals, obtained via simulated annealing algorithm. Detailed structure files can be found in the zip file attached.

| x=<br>n= | 1 | 2 | 3 | 4 | 5 | 6 | 7 | 8 | 9 | 10 |
|---|---|---|---|---|---|---|---|---|---|---|
| **0+x** | 3.35 | 6.05 | 8.50 | 10.29 | 12.09 | 13.64 | 15.44 | 16.84 | 18.39 | 19.94 |
| **10+x** | 21.09 | 22.64 | 23.79 | 24.44 | 24.94 | 26.49 | 28.03 | 29.18 | 30.73 | 31.63 |
| **20+x** | 32.53 | 33.03 | 34.58 | 35.23 | 36.78 | 37.93 | 38.58 | 39.08 | 40.63 | 41.53 |
| **30+x** | 42.18 | 42.68 | 44.23 | 45.13 | 45.77 | 46.27 | 47.82 | 48.62 | 49.12 | 49.62 |
| **40+x** | 51.17 | 52.07 | 52.72 | 53.22 | 54.12 | 54.62 | 56.17 | 56.57 | 57.47 | 57.97 |
| **50+x** | 59.52 | 60.17 | 60.82 | 61.32 | 61.97 | 62.47 | 64.02 | 64.66 | 65.11 | 65.61 |
| **60+x** | 66.11 | 66.61 | 67.11 | 67.61 | 68.11 | 69.66 | 70.46 | 70.96 | 71.46 | 73.01 |
| **70+x** | 73.81 | 74.31 | 74.81 | 75.46 | 75.96 | 77.51 | 78.16 | 78.61 | 79.11 | 79.61 |
| **80+x** | 80.11 | 80.61 | 81.11 | 81.61 | 83.16 | 83.85 | 84.35 | 84.60 | 85.10 | 85.60 |
| **90+x** | 87.15 | 87.95 | 88.45 | 88.95 | 89.25 | 89.75 | 90.25 | 90.75 | 91.25 | 92.80 |
| **100+x** | 93.60 | 93.90 | 94.40 | 94.65 | 95.15 | 95.65 | 96.15 | 96.65 | 98.20 | 99.00 |
| **110+x** | 99.30 | 99.80 | 100.05 | 100.55 | 101.05 | 101.55 | 102.05 | 103.59 | 104.39 | 104.69 |
| **120+x** | 105.19 | 105.44 | 105.94 | 106.19 | 106.69 | 107.19 | 108.74 | 109.54 | 110.04 | 110.34 |
| **130+x** | 110.84 | 110.99 | 111.49 | 111.99 | 112.49 | 112.99 | 113.49 | 113.99 | 115.54 | 116.14 |
| **140+x** | 116.39 | 116.89 | 117.14 | 117.64 | 118.14 | 118.64 | 119.14 | 120.69 | 121.28 | 121.78 |
| **150+x** | 121.93 | 122.43 | 122.68 | 123.18 | 123.43 | 123.93 | 124.43 | 124.93 | 125.43 | 126.98 |
| **160+x** | 127.58 | 128.08 | 128.23 | 128.58 | 128.88 | 129.23 | 129.63 | 129.88 | 129.88 | 130.38 |
| **170+x** | 130.88 | 131.38 | 131.88 | 132.38 | 132.88 | 134.43 | 135.03 | 135.28 | 135.78 | 136.03 |
| **180+x** | 136.53 | 137.03 | 137.53 | 138.03 | 139.57 | 140.17 | 140.67 | 140.82 | 141.32 | 141.57 |
| **190+x** | 142.07 | 142.32 | 142.82 | 143.32 | 143.82 | 144.32 | 145.87 | 146.37 | 146.87 | 147.12 |
| **200+x** | 147.47 | 147.77 | 148.12 | 148.52 | 148.77 | 148.77 | 149.27 | 149.77 | 150.27 | 150.77 |
| **210+x** | 151.27 | 151.77 | 153.32 | 153.57 | 154.07 | 154.32 | 154.82 | 155.07 | 155.57 | 155.82 |
| **220+x** | 156.32 | 156.57 | 157.07 | 157.57 | 159.11 | 159.86 | 160.21 | 160.51 | 160.86 | 161.26 |
| **230+x** | 161.51 | 161.51 | 162.01 | 162.26 | 162.76 | 163.01 | 163.51 | 164.01 | 164.51 | 165.01 |
| **240+x** | 166.56 | 167.06 | 167.41 | 167.71 | 168.06 | 168.46 | 168.71 | 168.71 | 169.21 | 169.46 |
| **250+x** | 169.96 | 170.21 | 170.71 | 171.21 | 171.71 | 172.21 | 173.76 | 174.25 | 174.60 | 174.90 |
| **260+x** | 175.25 | 175.65 | 175.90 | 175.90 | 176.40 | 176.65 | 177.15 | 177.40 | 177.90 | 178.40 |
| **270+x** | 178.90 | 179.40 | 180.95 | 181.45 | 181.70 | 182.10 | 182.35 | 182.35 | 182.85 | 183.10 |
| **280+x** | 183.60 | 183.85 | 184.35 | 184.60 | 185.10 | 185.35 | 185.85 | 186.35 | 187.90 | 188.55 |
| **290+x** | 188.90 | 189.30 | 189.55 | 189.55 | 190.05 | 190.30 | 190.80 | 190.95 | 191.30 | 191.70 |